# AMPA, NMDA AND GABA$_A$ RECEPTOR MEDIATED NETWORK BURST DYNAMICS IN CORTICAL CULTURES *IN VITRO*

*H. Teppola[1,2], S. Okujeni[2], M.-L. Linne[1], and U. Egert[2]*

[1] Department of Signal Processing, Tampere University of Technology, Tampere, Finland
[2] Bernstein Center Freiburg & Department of Microsystems Engineering, IMTEK, Albert-Ludwig University of Freiburg, Freiburg, Germany

heidi.teppola@tut.fi, okujeni@bcf.uni-freiburg.de, marja-leena.linne@tut.fi, egert@imtek.uni-freiburg.de

**ABSTRACT**
In this work we study the excitatory AMPA, and NMDA, and inhibitory GABA$_A$ receptor mediated dynamical changes in neuronal networks of neonatal rat cortex *in vitro*. Extracellular network-wide activity was recorded with 59 planar electrodes simultaneously under different pharmacological conditions. We analyzed the changes of overall network activity and network-wide burst frequency between baseline and AMPA receptor (AMPA-R) or NMDA receptor (NMDA-R) driven activity, as well as between the latter states and disinhibited activity. Additionally, spatiotemporal structures of pharmacologically modified bursts and recruitment of electrodes during the network bursts were studied. Our results show that AMPA-R and NMDA-R receptors have clearly distinct roles in network dynamics. AMPA-Rs are in greater charge to initiate network wide bursts. Therefore NMDA-Rs maintain the already initiated activity. GABA$_A$ receptors (GABA$_A$-Rs) inhibit AMPA-R driven network activity more strongly than NMDA-R driven activity during the bursts.

## 1. INTRODUCTION

Synchronous patterns of activity, accompanied with intracellular Ca$^{2+}$-transients, are considered to play an essential role in the development of neuronal networks in a wide range of brain structures [4], including networks of dissociated cortical neurons *in vitro* [3]. Synchronous activity in form of network-wide bursts (NB) is thought to be generated by recurrent excitatory pathways. The inhibitory pathways suppressing excitation are also considered important in network burst dynamics.

As in native cortical tissue, glutamatergic AMPA-Rs (expressing fast ion channel kinetics) and NMDA-Rs (expressing slow ion channel kinetics, and both voltage and Mg$^{2+}$-dependency) are the main mediators of excitatory synaptic transmission among neurons *in vitro*. Fast inhibition is mediated through GABAergic transmission via GABA$_A$-Rs. Despite the solid biophysical characterization of AMPA, NMDA, and GABA$_A$ receptors at the monosynaptic level, their complex interplay on the network level is still not fully understood.

Simultaneous multi-unit recordings from neuronal networks of an intact brain *in vivo* are challenging to conduct with the existing technology. Therefore, microelectrode array (MEA) measurement systems have been developed for simultaneous extracellular recordings of many neurons in *in vitro* networks. These recordings allow spontaneous and stimulus evoked NB activity to be recorded over hours of time [5]. Previous spontaneous recordings of neural activity at 21 days *in vitro* (DIV) with MEA have shown, similarly to this work, that networks posses stable dynamics following synapse formation and maturation. The characteristic network-wide bursting in these previous studies consists of around 0.5 to 1 second long periods of simultaneous high spiking activity on many electrodes followed by 0.5 to 20 seconds quiescence periods [3], [6], [7].

Pharmacological studies of stimulus evoked network activity in cortical cultures have shown that the late component of NBs (25ms after stimulus, lasting several hundred milliseconds) was reduced or abolished by NMDA-R antagonist, such as APV and high Mg$^{2+}$, without affecting to the early component (first 25ms after stimulus). When inhibitory GABA$_A$-Rs were blocked by bicuculline (BIC) the opposite effect was obtained, the late component of evoked responses was increased [1]. Another study in dissociated spinal cord cultures has shown that the duration of AMPA-R driven bursts decreased compared to the baseline, suggesting the role of NMDA-Rs in the maintenance of high network activity during bursting. In addition, NMDA-R antagonist has been found to reduce the burst rate indicating a contribution of NMDA-Rs in network activation. By silencing AMPA-R at the presence of disinhibition (BIC), the burst rate was reduced and burst onset was slowed down and in some experiments bursting ceased completely by AMPA-R antagonist [2].

In this work, we study for first time the coordinated interplay between excitation and inhibition in network dynamics in dissociated neurons of rat neonatal cortex. First, we obtained spontaneous baseline activity with MEA from six different densely cultured networks at 21 DIV (**Figure 1**). Then, we silenced each of the excitatory pathways by blocking either AMPA-Rs or NMDA-Rs with NBQX or D-AP5, respectively, both in three different cultures. Finally, we disinhibited the networks with $GABA_A$-R antagonist PTX. This protocol allows us to investigate the temporal and spatial patterns of network dynamics, produced by fast and slow excitatory and fast inhibitory transmission.

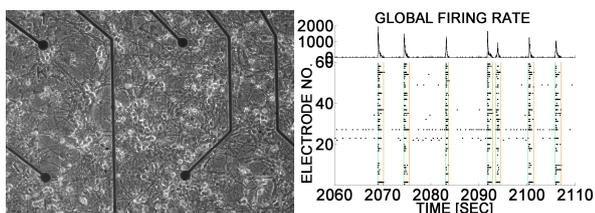

**Figure 1. Left:** Cultured network of cortical neurons on MEA plate at 21 DIV. **Right:** Spontaneous synchronized patterns of activity. The upper plot shows array wide firing rate (Hz) and the raster plot at the bottom shows the activity (spikes) recorded with 59 electrodes during a time period of 50 seconds.

## 2. MATERIALS AND METHODS

**Primary cell cultures:** Cortical cells were extracted from neonatal Wistar rats' frontal lobes and enzymatically and mechanically dissociated. 200 000 cells were seeded on 0.1% polyethyleneimine coated microelectrode arrays (60 TiN electrodes Ø 30µM, 500µM electrode pitch, Multi Channel Systems Ltd.). Cells were cultured in 2ml minimum essential medium supplemented with 5% heat inactivated horse serum, 20mmol/l glucose, 1% gentamycine and 0.5mmol/l L-glutamine in 5% $CO_2$ humidified incubator at 37°C. One third of the medium were changed twice a week and recordings were done between 21-23 DIV.

**Pharmacology:** The following agents were used for suppressing ionotropic receptors: 10µM D-(-)-2-Amino-5-phosphonopentanoic acid (**D-AP5**) to competitively block the glutamate site of NMDA-Rs, 10µM 2.3-Dioxo-6-nitro-1,2,3,4 tetrahydrobenzo-[f]quinoxaline-7-sulfonamide (**NBQX**) to competitively block AMPA-Rs and 10µM Picrotoxin (**PTX**) to non-competitively block $GABA_A$-R. All drugs were applied to medium with a pipette.

**Electrophysiological recordings and analysis:** Spontaneous network activity was recorded with MEA-1060 amplifier (Multi Channel Systems Ltd.) in normal culture conditions and after applying each of the drugs for one hour. Raw signals from each electrode were digitally high-pass filtered at 200Hz and extracellular activity of neurons (spikes) were detected based on voltage threshold, -5 times of standard deviation from mean noise baseline of each electrode using MC-Rack software (Multi Channel Systems Ltd.).

For further analysis the initial 600s of recorded data after drug application were discarded and the following 3000s were analyzed. Artifacts and electrodes with firing rates (FR) lower than 9% of the average FR on electrodes with spike activity were removed. NBs were detected according to following criteria: first inter spike interval < 100ms defines onset time of NB and first inter spike interval > 100ms defines offset of NB, in addition, minimum 8 electrodes should be active during the NB.

## 3. RESULTS

We pharmacologically blocked the excitatory AMPA-R and NMDA-R separately. Upon monitoring network activity, we additionally blocked $GABA_A$-R to observe the interplay between inhibition and each excitatory pathway.

First, we calculated the overall network activity (spikes/sec) and burst frequency (bursts/min) over whole analysis period of 3000s. Recordings from networks under AMPA-R blockage (n=3) and under NMDA-R blockage (n=3) were compared to the baseline recordings. As expected, overall firing rate decreased upon suppression of either AMPA-R or NMDA-R driven excitation and increased when networks were disinhibited as shown in **Table1**. When comparing the ratio of overall firing rate and burst frequency between baseline and NMDA-R suppressed excitation the results showed that the overall firing rate and burst frequency decreased by a factor of 10 and 2, respectively. However, blocking AMPA-Rs decreased overall firing rate only 3 fold and burst frequency 21 fold (**Table1**).

When the excitatory pathway was driven by AMPA-Rs and the inhibitory pathway was simultaneously suppressed, the overall network activity increased 14 fold (n=3) but burst frequency decreased 2 fold (n=2) in two out of three cultures and increased 3 fold (n=1) in one out of three cultures. Disinhibition during NMDA-R driven activity caused overall network activity to increase only 2 fold and burst frequency to increase 5 fold (**Table1**).

Burst firing rate profiles in **Figure 2** show how the temporal structure of bursts are modified following each drug application at single electrode level and at average over all electrodes. AMPA-R driven bursts became shorter, contained fewer spikes, the peak FR was reached earlier (~20-40 ms) in the bursts and late phase was diminished. Contrarily, NMDA-R driven bursts became longer, contained more spikes, the peak FR was achieved later in the course of the burst (~ 90-200 ms) and the late phase of NBs was more pronounced. Disinhibition increased the early and late phases of bursts in comparison to control and AMPA-R driven bursts but, interestingly, in comparison to NMDA-R driven bursts the late phase squeezed after disinhibition **(Figure 2)**.

The recruitment of electrodes during NBs was significantly decreased during AMPA-R driven activity. In contrast, networks were widely recruited over 200ms time periods with disinhibition of AMPA-R driven activity as shown in **Figure 3**. Approximately the same number of electrodes was recruited when NMDA-R driven NBs were compared to control case. However, after disinhibition, the recruitment of electrodes a bit increased but more importantly recruitment was much faster than during NMDA-R driven or baseline activity, indicating that NMDA-R driven inhibition slows down the propagation of network-wide activity at the beginning of bursts **(Figure 3)**.

## 4. CONCLUSIONS

Our results show that AMPA-R driven NBs are initiated more often than NMDA-R driven ones contributing over ten times more to the initiation of network activity compared to NMDA-Rs. By blocking NMDA-Rs, the overall firing rate was lowered three times more compared to AMPA-R blockage, which indicates the importance of NMDA-Rs in maintaining network activity during the bursts.

Nearly ten times higher overall firing rates were observed during disinhibited AMPA-R driven activity compared to the NMDA-R driven one. This indicates the $GABA_A$-Rs primary role for inhibiting especially AMPA-R driven network activity. The connection between the roles of $GABA_A$-Rs and AMPA-Rs is also indicated by the fact that NMDA-R driven activity displays higher firing rates compared to the control case, where both NMDA-Rs and AMPA-R are active **(Figure 2)**.

Contradictory behavior of both decreasing and increasing burst frequencies during disinhibition of AMPA-R driven activity could be explained by two different underlying mechanisms: 1. There are more bursts, due to lower firing threshold, to initiate AMPA-R driven NBs when inhibition is suppressed; 2. There are less bursts because AMPA-R driven neurons are more excited, thus lengthening network bursts and, hence, recovery times.

We additionally found out that disinhibition, in general, increases the early and late phase of bursts due to decreased threshold for excitation. Suprisingly, the late phase of NMDA driven bursts shortened and burst frequency increased by disinhibition. This can be explained by our finding that $GABA_A$-R slows down the recruitment of NMDA driven NBs, thus by blocking $GABA_A$-Rs in such a network, the recruitment of NBs speeds up and further, the duration of bursts shortens.

Our results further indicate that the dynamics of cortical networks *in vitro* incorporate a complex interplay between excitatory and inhibitory transmission. NMDA-R and AMPA-R driven activities are both able to initiate network-wide bursting. However, AMPA-Rs role for initiation is much more important. The main role of NMDA-Rs is to maintain already initiated activity. $GABA_A$-Rs inhibit AMPA-R driven network excitation more strongly than the NMDA-R driven during bursts. However, $GABA_A$-R mediated inhibition slows down the propagation of NMDA-R mediated network activity more than AMPA-R mediated. Experimentally investigated interplay between excitatory and inhibitory transmission in cortical networks *in vitro* can further be studied using biologically plausible computational network models.

## 5. ACKNOWLEDGEMENTS

The authors would like to acknowledge the following funding: Graduate School of Tampere University of Technology, Tampere Graduate School in Information Science and Engineering (TISE), Academy of Finland Grant (213462, 129657, 106030, and 124615), German BMBF (FKZ 01GQ0830, 01GQ0420), and EC (NEURO, 12788).

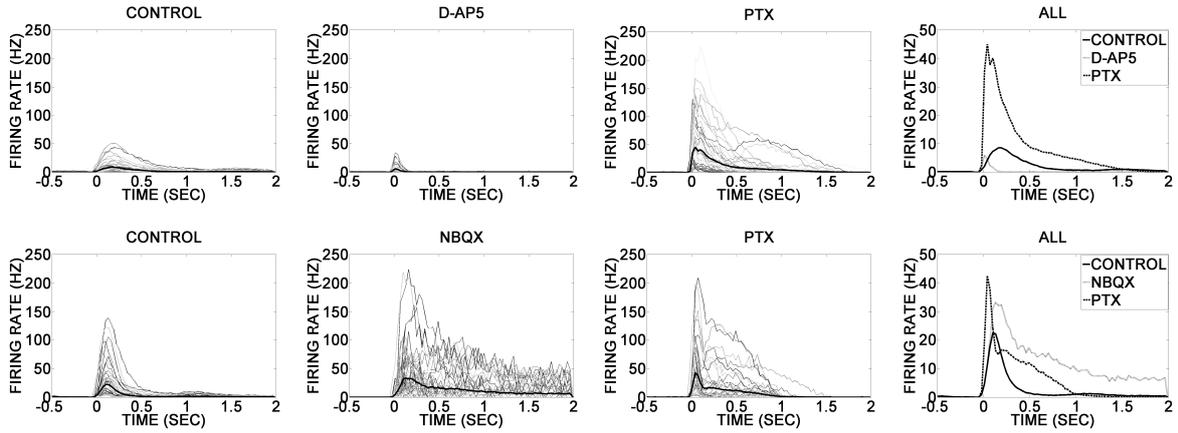

**Figure 2. Burst firing rate profiles.** Upper row corresponds to AMPA driven experiment and lower row to NMDA driven experiment. Columns from left: firing rate profiles during control, suppression of excitation, suppression of inhibition, and the average profiles for all three cases. D-AP5, NBQX, and PTX are the pharmacological agents for suppressing NMDA-Rs, AMPA-Rs, and GABA$_A$-Rs respectively. In the first three columns one single thin line corresponds to firing rate in one electrode averaged over all bursts (aligned to NB onsets). In all columns, thick lines correspond to overall average burst firing rates over all electrodes and over all bursts (bin width 30ms).

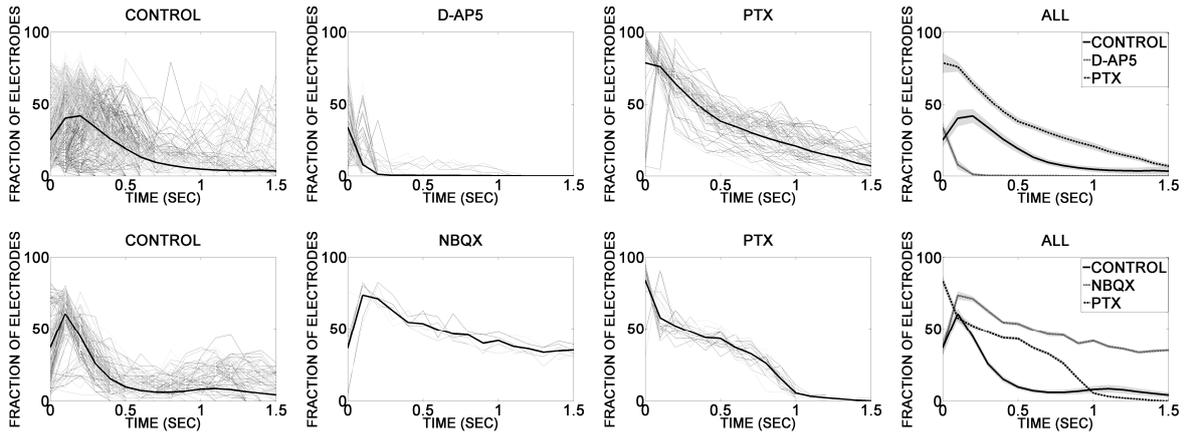

**Figure 3. Network burst electrode fraction curves.** Upper row corresponds to AMPA driven experiment and lower row to NMDA driven experiment. Columns from left: Percentage of the electrodes attending to the NBs for control, suppression of excitation, suppression of inhibition, and the average curves for all three cases. D-AP5, NBQX, and PTX are the pharmacological agents for suppressing NMDA-Rs, AMPA-Rs, and GABA$_A$-Rs respectively. In the first three columns, one single thin line corresponds to fraction of attending electrodes in one network burst. In all columns, thick lines correspond to average fraction of electrodes over all network bursts. In the most right column, shaded areas show the standard error of mean (bin width 100ms).

| Pharmacological condition | AMPA driven activity | NMDA driven activity | Disinhibition for AMPA driven activity | Disinhibition for NMDA driven activity |
|---|---|---|---|---|
| Increase in overall firing rate | — | — | 14.1±2.3(n=3) | 1.8±0.5 (n=3) |
| Decrease in overall firing rate | 9.8±2.9 (n=3) | 3.2±1.7 (n=3) | — | — |
| Increase in burst frequency | — | — | 3.2 (n=1) | 4.6±2.0 (n=3) |
| Decrease in burst frequency | 2.2±0.7 (n=3) | 21.3±9.2(n=3) | 2.4±1.3 (n=2) | — |

**Table1.** Computed changes in overall firing rate (spikes/s) and burst frequency (bursts/min) compared to the previous pharmacological condition (mean ± standard deviation). Disinhibition chronologically follows the associated AMPA-R driven or NMDA-R driven activity.